# Secure Quantum Computing for Healthcare Sector: A Short Analysis

P. Srikanth[a], Adarsh Kumar[b1]

[a,b] Systemics Cluster, School of Computer Science, University of Petroleum and Energy Studies, Dehradun, Uttrakhand-248001, India.
*pshrikant@ddn.upes.ac.in,adarsh.kumar@ddn.upes.ac.in.*

**Abstract**

Quantum computing research might lead to "quantum leaps," and it could have unanticipated repercussions in the medical field. This technique has the potential to be used in a broad range of contexts, some of which include the development of novel drugs, the individualization of medical treatments, and the speeding of DNA sequencing. This work has assembled a list of the numerous methodologies presently employed in quantum medicine and other disciplines pertaining to healthcare. This work has created a list of the most critical concerns that need to be addressed before the broad use of quantum computing can be realized. In addition, this work investigates in detail the ways in which potential future applications of quantum computing might compromise the safety of healthcare delivery systems from the perspective of the medical industry and the patient-centric healthcare system. The primary objective of this investigation into quantum cryptography is to locate any potential flaws in the cryptographic protocols and strategies that have only very recently been the focus of scrutiny from academic research community members.

## 1. Introduction

Quantum computing (QC) leverages quantum mechanics with the physical quantum theory, which includes quantum superposition and entanglement [1]. The QC employs the qubits or quantum bits, whereas traditional computing uses the bits. As a result, QC processes a massive amount of data parallelism in real-time over the qubits-based computing environment [2]. Thus, QC encourages researchers to employ a variety of areas like image processing, cryptography, electronics, information, and communication theory [3], [4]. However, the QC has resulted in substantial revolutions in various domains, including financial modeling, physics, cryptography, transportation, and weather forecasting. Further, the QC is ideally suited for several computer-intensive applications such as healthcare [5]. The digital healthcare framework is interconnected with medical devices through the internet to the cloud [6], [7], as illustrated in Fig.1.

According to Fig.1, the existing healthcare communication network is classified into three layers: perception, internet, and cloud or cellular network. The framework's primary purpose is to deliver high quality of services (QoS) to healthcare applications through medical devices. The medical devices are placed in the perception layer, such as healthcare infrastructure, medical sensors, machinery, patients, practitioners, and health workers [8]. These devices have constrained processing power which can be resolved by integrating with high infrastructure platforms such as cloud computing environment, cellular network, or many others that are included as the third layer. Further, medical devices suffer from substantial connectivity issues with the actuators and sensors. Hence, the perception layer establishes short-range communication by leveraging various protocols such as ZigBee, 6LoWPAN, Wi-Fi, and Bluetooth, and these communication protocols are connected to a cloud/cellular network. However, the sensors and actuators are third-party standardized cellular solutions that leverage the license-exempt spectrum; thus, there is no guarantee of QoS. Therefore, internet-based healthcare applications require QoS and massive computational processes. Therefore, smart healthcare devices deliver healthcare services everywhere at any time by connecting to the internet and providing the QoS by collaborating in real-time [9], [10]. Although the computational speed of classical computing needs to be increased, healthcare applications offer more effective and efficient solutions to patients. Hence, adopting the QC permits massive data processing that enables patient-centric treatment and diagnosis. Further, the hospital infrastructure is brought to the cloud environment that predicts diseases faster because of qubit processing.

[*] Corresponding author. Tel.: +91-9557654451.
*E-mail address:* pshrikant@ddn.upes.ac.in

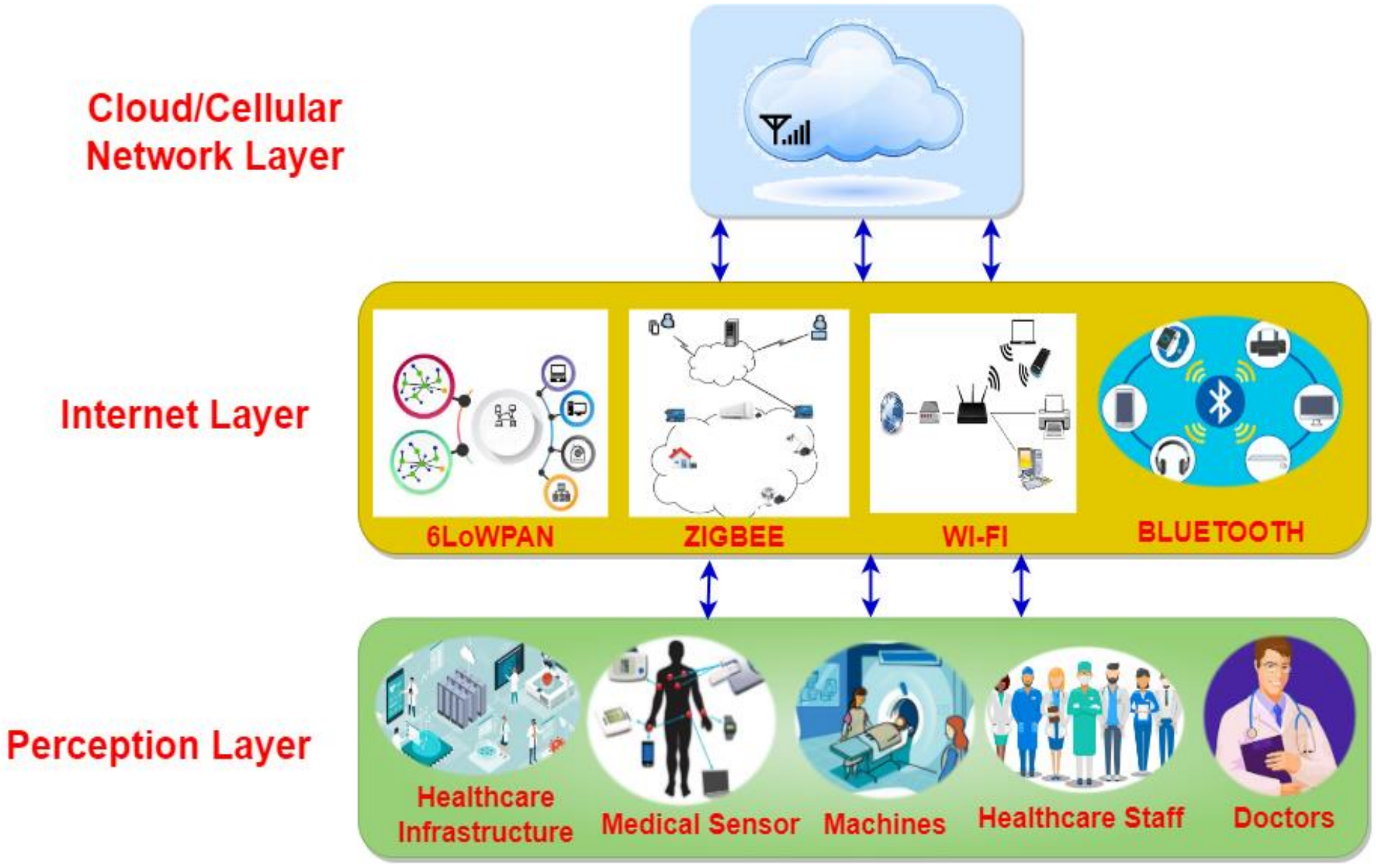

Fig. 1 Healthcare Communication Network

In this work, the healthcare communication network is discussed briefly before presenting the importance of quantum computing in the healthcare sector. Here, the application of quantum computing in healthcare and its importance in future are discussed. This discussion is developed by identifying the important factor that can create impacts in the healthcare domain. In comparative literature analysis, various healthcare security schemes are analyzed especially those schemes identified which have discussed the importance of quantum computing. Further, software tools, techniques and languages are discussed that make quantum computing special for healthcare and other application domains. This also shows the speed and volume of growth of quantum computing in recent times. Finally, open issues, challenges and future directions of integrating quantum computing with healthcare are explored. The contributions of work (briefly) are as follows:

- The healthcare paradigm applications and challenges associated with stakeholders like hospitals, patients, pharmaceutical organizations, and insurance organizations are demonstrated. Further briefly explained the adoption of QC to overcome the difficulties of traditional computing.
- The state-of-the-art of quantum-based healthcare paradigm security challenges are demonstrated
- Several quantum-based simulation tools and programming languages utilized by simulation tools are demonstrated.
- The different outstanding challenges and potential future directions for the healthcare paradigm are described.

The article is streamlined as section II presents the quantum computing for healthcare paradigm, section III describes the security of QC healthcare, and section IV outlines the various simulation tools and languages available for qubit processing. Section V presents the open challenges and future directions toward the healthcare paradigm, and Section VI describes the conclusion.

## 2. Quantum computing for healthcare

The QC enables significant improvements in computational power, resulting in healthcare industry improvements. The QC emphasizes the numerous innovations in healthcare applications listed in Fig. 2. The QC permits supersonic drug formulation and silico diagnostic testing through virtual humans. Deoxyribonucleic acid (DNA) sequencing with QC offers a personalized treatment. The innovative therapy and drugs are developed through comprehensive modeling. The QC tackles complex optimization problems like effective radiation plans to eliminate the designated cancer cells while avoiding damage to healthy organs and body parts [11]–[13]. The atomic-level molecular interaction analysis permits drug development and medical research. The genome sequencing and analysis employing qubits is accomplished in a short period. Further, hospital infrastructure has been migrated to the cloud, predicting chronic medical conditions and securing medical records through qubit processing. The introduction of QC in healthcare paradigms provides exponential benefits such as improving patient management, promoting medical professional experiences, lowering costs, and delivering better patient treatment [14].

The healthcare sector data has progressed enormously, including clinical studies, illness repositories, electric health records (EHR), and medical device inspections. As a result, healthcare supervisors must make real-time decisions using the extracted data by leveraging complex systems. Thus, recent research uses QC, which provides various benefits. Likewise enhances the speed of illness diagnosis and treatment, computational speed into minutes, IT architectures, and corporate strategies based on the application. Further, high-level data privacy for healthcare is required by the QC features that enable a variety of use cases such as accelerating diagnoses, personalized treatment and price optimization. Moreover, the adoption of QC enhances access to healthcare-related data. However, the differences between traditional computing (TC) and QC are computational capacity of QC increases exponentially, whereas it linearly increases in TC. The computing units in TC are binary values (0 or 1); QC qubits are used. The environment with the sustainability of TC is room temperate with non-compute intensive, whereas QC ultra-cold is required with compute-intensive. The processing unit is the CPU in TC, quantum processing unit (QPU) in QC. The error rate is low in TC, and the governance is classical physics, whereas the error rate is high, and quantum physics or mechanism is used in QC to govern the process. Further, the building block in TC is a CMOS transistor, whereas superconducting quantum interference is used in QC [15]. Therefore, the healthcare paradigm is adopted the QC because of the characteristics mentioned above that enable quick, personalized, and precise healthcare services.

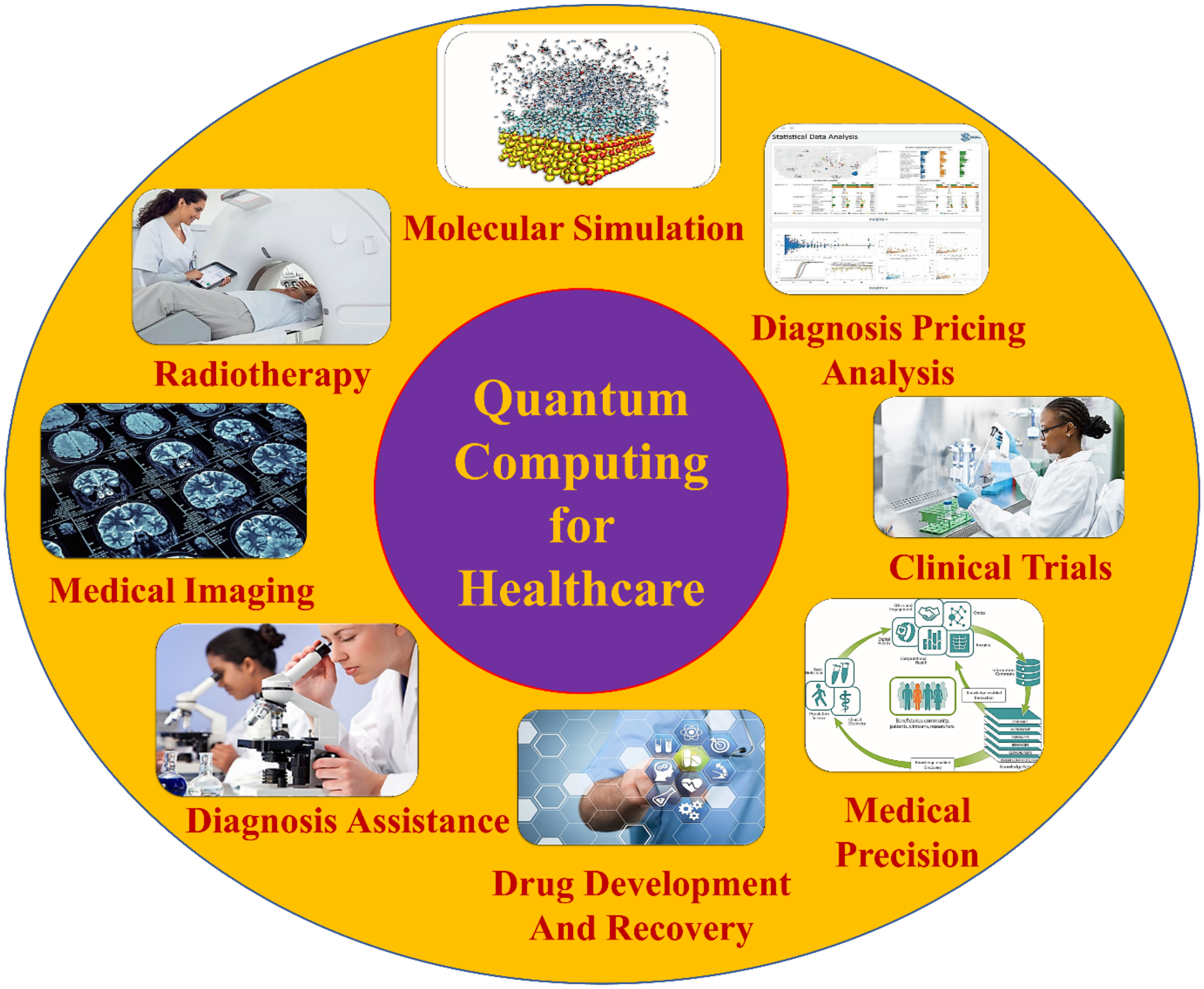

Fig. 2 Quantum-based innovation in healthcare applications

According to Fig.2, the healthcare-related applications are demonstrated as

- *Molecular Simulations:* The quantum-based healthcare systems leverage Artificial Intelligence (AI), machine learning (ML), deep learning (DL), and optimization technologies employed in computationally complex problems such as molecular simulations [16]. The molecular simulations are highly coupled components such as chemical compounds with multiple protons that are interrelated and offer efficient healthcare processing [17]. Therefore, traditional computing challenges such as computational power and resource requirements are easily managed through QC.

- *Diagnosis pricing analysis:* The QC improves the risk analysis by forecasting the patient's present health status by expecting the patient's tendency to impact a specific illness. As a result, insurance pricing and rates can be optimized. In conjunction with quantum risk models, the illness risk examination at the species scale will assist the computation of the pricing and financial risks at the granular stage. Further, the detection of healthcare fraud avoids revenue loss. The QC enhances the accuracy of the classification and malicious behavior based on pattern recognition over the traditional approaches in medical claims [14].

- *Drug development and recovery:* The QC enables healthcare professionals to perform medical research at the atomic level in complex molecular interactions [17]. Thus, it permits the medical professional to research various aspects, including diagnosis, drug discovery, therapy, and analysis. The advancement in the QC enables the interactions with proteins in the human genome can be simulated. However, AI approaches are enormously being used to help with patient diagnosis. The pattern recognition based on the ML scheme is trained on large-scale data and validates the current data with training data that can produce accurate treatment and diagnosis. Hence, the QC will be helpful for health staff to make quick decisions according to the appropriate pattern.

- *Medical precision:* The objective of medical precision is to offer patient centric treatment according to an individual patient's disease. Therefore, in the future, patient-centric medical treatment will be essential for the complex biological system of human beings. However, the medicine cost is associated with 10-20% of healthcare contributions, and 80-90% are related to other expenses, including socio-economic factors, environmental factors, and health behavior challenges. Further, drug-based therapies cannot obtain effective results for individuals that may lead to deaths due to drug reactions. Therefore, early treatment and prevention interferences can enhance healthcare objectives and lower costs through QC [14]. Predicting future disease risk assessment using EHR is effectively performed through traditional schemes, but some level of noise, quality, size, and complexity are the challenges of conventional systems. Therefore, quantum-based ML enhances the precision of early disease discovery. Further, healthcare practitioners can easily discover the disease by leveraging medical devices that mitigate the risk through continuous monitoring and provides the appropriate treatment. Therefore, quantum ML can enhance patient-centric treatment through continuous data streaming through medical devices, which offers continuous services to patients.

- *Diagnosis assistance:* It is used to provide an early stage of disease; as a result, it reduces the healthcare cost and delivers enhanced diagnosis and treatment [14]. For instance, early-stage detection of the cancer and COVID-19 will decrease the treatment cost significantly. The diagnosis tool such as X-rays, MRIs, and CT scans are expensive computer-aided devices that are developing faster [18]. However, these devices suffer from critical challenges of noise, quality, replicability, and safety. Therefore, QC aids diagnosis by examining the medical images through edge detection, enhancing the medical-aided diagnosis. Furthermore, categorizing cells based on biochemical and physical features requires ample space because predictor variables are more [19]. These critical challenges can be resolved through quantum-enhanced ML techniques such as quantum vector space that enhances the single-cell diagnosis. The adaption of QC avoids the repetitive diagnosis and treatment and offers regular monitoring of individual health.

- *Radiotherapy:* It is used for cancer treatment by leveraging electromagnetic energy to destroy malignant cells and prevent their proliferation[20]. However, radiation is a delicate technique that requires rigorous computations to effectively target the disease while not harming healthy organs in the body. The radiotherapy is carried out by leveraging exceedingly accurate devices which involve a high precision optimization problem. Therefore, achieving the precise radiography procedure requires several accurate and sophisticated simulations to obtain an effective solution. The adoption of QC enables various simulation recommendations that permit the numerous simulations to occur concurrently and to determine an effective solution faster [21].

- *Medical imaging:* The healthcare industry is revolutionized with many aspects, such as medical imaging, which permits clinicians and health practitioners to identify disease early and provides treatment to patients that help them faster recovery[22]. The medical imaging assists healthcare practitioners, insurers, and patients in understanding the body structure and making real-time decisions. Further, medical imaging requires enhancement when the complex parameters are identified in body structure, then extracting the key features through QC-based machine learning or artificial intelligence enables quick processing for medical imaging.

- *Clinical trials:* Clinical study on coronavirus and its variations, as well as other illnesses like AIDS and TB, can help find the machinery and tools needed to build countermeasures against chronic and other complicated diseases. The coronavirus is a component of the SARS virus, which poses a major risk to humans today[23]. Pharmaceutical firms have created a few vaccinations, however no vaccines for children aged 0 to 18 have been developed. Clinical trials are one method of verifying and strengthening the dependability of a medicine or vaccination. In most clinical studies, animals or people are used as testing carriers, which may be avoided using silico tests, in which quantum computers can imitate human beings[24]

The challenges in healthcare applications based on traditional computing are overwhelmed through the QC adoption summarized in Table1.

Table 1: Summary of the healthcare applications

| Stakeholder | Application | Challenge | Opportunity with the QC |
|---|---|---|---|
| Hospitals | Medical imaging [22] | Medical imaging requires various image processing techniques to collect the information from the complex body structure that requires more processing. The TC is inefficient in terms of speed and accuracy. | The qubit processing environment speeds up the computational capacity and provides accurate results for various image processing that helps clinicians make real-time decisions. |
| | Radiotherapy[20] | Complex diseases like cancer treatment require radiotherapy that can be done without harming the healthy body cells. Hence it requires large-scale optimization, high computational power, and enhanced accuracy. | The QC offers large-scale optimization that provides enhanced convergence rates and effective radiotherapy. However, radiotherapy is associated with most surgeries. Hence, QC improves the accuracy and supplies high precision for treatment because of the high computation power. |
| | Diagnosis assistance[18] | The existing diagnosis approaches are complex, complicated, and expensive. Moreover, the wrong diagnosis increases the cost of the treatment and repeated diagnosis. | The QC adoption reduces the diagnosis cost by eliminating repetitive tests by leveraging quantum machine learning and quantum vector space. Further, it offers regular monitoring of individual health. |
| Patients | Medical precision [14] | The patient is treated according to the type of disease based on the doctor's experience.<br>The sensors and actuators collect the patients' data and perform various operational and clinical activities. Therefore, the data generated need to be maintained safe and reliable, which enhances the patient experience.<br>The diagnosis treatment requires quick solutions according to diseases. | The QC determines the big data through the sensors and actuators connected to high infrastructure networks, i.e., the cloud, and performs the regular monitoring of the individual.<br>The quantum-based cryptograph algorithms provide more security compared to the traditional algorithms that enhance the safety and reliability of data.<br>Through employing the QC, patient-centric treatment and disease-based drug suggestions |
| | | | are possible. Hence it provides quick solutions. |
| Pharmaceutical organizations | Drug development and recovery [17] | The classical approach consumes more time to develop the drugs and vaccines based on the disease type. | The quantum-based AI approaches are enormously used at the atomic level of complex molecular interactions. Thus it permits medical professionals to do their research in various aspects, including diagnosis, drug discovery, therapy, and analysis quicker. |
| | Molecular Simulation[16] | The molecular simulation requires many resources that increase in size exponentially. Further, it requires a high processing speed. | The QC manages the processing speed with qubits. Further, quantum-based AI and ML techniques are helpful for complex simulations. |
| | Clinical trials[23],[24] | Clinical trials test the strength and reliability of therapy or vaccine on animals or humans. | The qubit processing can verifies the accuracy and precision of therapy or vaccine in simulating human beings |
| Insurance organizations | Diagnosis pricing analysis [17] | The current health insurance companies are struggling with medical claims fraud. | Healthcare fraud can be optimized by using QC-based classification and malicious behavior pattern recognition in medical claims. |

Although the expansion of QC provides innovative opportunities in the pharmaceutical industry, it is essential for the healthcare paradigm. However, the adoption of quantum methods poses security concerns in healthcare services and the pharmaceutical industry. The healthcare organization relies on web-based data exchange by providing services to connected healthcare devices. The healthcare devices acquire services that can be attacked in numerous manners [25]–[27]. The pharmaceutical sector is suffering from poor drug quality, leading to millions of fatalities yearly. Hence, it requires security challenges such as traceability that finds the medicine suppliers who supplied the expired medicines. Further, the healthcare infrastructure is plagued by security problems. For instance, the most extensively employed public key cryptography techniques provide exceptional security in communication networks. However, computational and communication technology breakthroughs have effortlessly demolished the endeavors. As a result, it necessitates the usage of the optimal implied key size [28]. Therefore, this study signifies the essence of designing a secure and privacy-preserved protocol for delivery services to medical devices by leveraging the QC.

## 3. State of the art

Numerous authors present QC-based techniques in the healthcare domain. The literature has briefly demonstrated recent work on the security of QC healthcare because the data is collected from medical devices and transmitted over the open channel such as the internet. Thus, the hackers can miss using sensitive data such as tampering or altering data, illegal distribution, creating security and privacy challenges. Hence, data protection and privacy protocols are essential in the quantum-based healthcare paradigm. Further, healthcare 4.0 leverages the internet of things (IoT) and cloud services to access medical data remotely with the consideration of healthcare 4.0 elements. Thus, this section describes the various security and privacy challenges of quantum healthcare. Childs et al. [29] described a cloud-based quantum computer that employs natural language processing approaches and collects geriatric patient healthcare information. This study aims to enable the evidence-based co-optimized diagnosis for geriatric patients in rural and urban areas. Datta et al. [30] proposed an aptamer for detection and diagnosis tool (ADD).The ADD is a smartphone application that collects data from entangled quantum nanodots and detects the molecules in the SARS-CoV2 virus. Although, Child [29] and Datta [30] approach are acquiring and storing the patients' data from healthcare devices. Since these approaches provide limited scalability and not focusing on patient data security and privacy. Janani et al. [31] developed a telehealthcare paradigm based on quantum-based spatial domain image encryption, providing two-level security for medical images. The developed model uses the cancer images fortified by the original seed values for encryption. Moreover, this model employs the novel approach of a quantum block-based scrambling that enables privacy for medical images. It also ensures data integrity through quantum encryption for data based on the region of interest (ROI). The proposed model is robust to statistical and differential attacks but has a high computational overhead during quantum cryptography. Latif et al. [32] proposed an encryption mechanism for healthcare patient data protection through controlled quantum walks. The quantum walks are computed with substitution and permutation. However, this approach provides robust results to protect the patients' privacy through image encryption but provides overhead. Bhavin et al. [7] proposed a secure electronic health record with quantum encryption and blockchain technology. The blockchain technology permits users to access them based on their privileges. Further, a quantum blind signature is leveraged to resist the quantum attacks while creating the blocks through the Hyperledger. Therefore, the proposed approach provides efficient throughput, network congestion, and resource consumption results. However, this approach efficiently works for a homogenous environment but still needs to test for a heterogeneous environment. Further, this approach provides network overhead because of the QC and blockchain infrastructure. Devi et al. [6] propose an effective health monitoring of patients remotely with social distance through a wireless body sensor network. The sensed information is communicated over the sensor network to the doctor, and the intruders can acquire the data by performing various attacks that cause various security threats. Hence, Devi [6] proposed a key distribution approach as an enhanced BB84 quantum cryptography protocol to share the key between the communication parties using quantum theory with bitwise operators. The experimental results demonstrate the proposed approach provides high security for various attacks, including wormhole, quantum, spoofing, blockhole, and Dos attacks. However, this approach needs to examine for coherent and individual eavesdropping attacks. Perumal et al. [28] proposed an optimized quantum key management technique for heterogeneous healthcare devices. The communication between authorized parties is established through quantum channels, and the generated key is distributed in terms of qubits via dedicated quantum channels. The key server generates the encryption key and sends it over the quantum channel. Then the content server encrypts the healthcare data and sends it over the public channel to healthcare practitioners. However this approach optimizes 90% of eavesdropping, but this approach suffers the communication overhead. Gong et al. [33] proposed a quantum k-means algorithm in quantum cloud computing. The quantum k-means execute the swap test and Grover optimization, reducing the clients' load. However, this approach allows regular clients to communicate with the quantum cloud server with small quantum machines. Azzaoui et al. [34] present the quantum cloud as a service for complex smart healthcare computations by leveraging the quantum terminal machines and blockchain. The healthcare data is protected through the Grover searching approach over the encryption on the quantum cloud with unconditional Q-OTP. However, this approach is examined under the medical image dataset, but it is essential to examine it under real-time medical data for determining the effective quantum error rate. Chen et al. [35] proposed to secure electronic medical records (EMR) through an Anti-quantum attribute-based signature (AQ-ABS) with blockchain technology that resists quantum attacks in healthcare. The proposed approach works based on the buyer-seller scheme, and their signatures are verified through AQ-ABS then the EMR data is stored in the interplanetary file system (IPFS).Therefore, this scheme provides anonymity, tamperproof, and control access. Mirtskhulava et al. [36] proposed a post-quantum digital signature and blockchain technology for the mobile internet of things (MIoT). Thus, this scheme speed-up the key generation process and provides security and privacy for EHRs. Further, blockchain technology resolves the various IoT device challenges in 5G and 6G networks. Mahajan et al. [37] and Fernandez-carames et al. [38] presented a study of blockchain-based solutions to provide robust data storage and sharing of healthcare data. Gupta et al. [39] designed a certificate-less data authentication scheme that enables the security characteristics of the internet of vehicles. The protocol features are to resist quantum attacks through lattice cryptography, provide trustworthiness in batch verification for vehicles with blockchain technology, and resist the forge-ability against selected message attacks. Cai et.al[40] developed a light-weight quantum blind signature with smart contract that improves the security against the quantum attacks. Singh et.al [41] provided a comprehensive survey on the healthcare data security approaches such as biometrics, watermarking, cryptography, and blockchain-related techniques. Karunarathne et.al examine the various security and privacy of IoT in the healthcare paradigm and suggested the security and privacy perspective solutions. Abdu et.al [42] proposed a grid based blockchain to manage the surgical process related data. The surgical records requires the robust record storage by integrating quantum of records. Therefore, this framework optimizes the transaction query process.

Table 2: The study of various healthcare security schemes

| Author | QC-healthcare | Security | IoT | Key feature | Findings |
|---|---|---|---|---|---|
| Childs [29] | √ | × | × | Cloud-based quantum computing | ➢ Limited scalability |
| Datta [30] | √ | × | √ | Smartphone app | ➢ Limited scalability |
| Janani [31] | √ | √ | √ | IPFS | ➢ Computational overhead |
| Latif [32] | √ | √ | × | Quantum walks | ➢ Robust to protect the patients' data<br>➢ Computational overhead |
| Bhavin [7] | √ | √ | √ | Blockchain technology | ➢ Network overhead |
| Devi [6] | √ | √ | √ | Quantum cryptography protocol | ➢ The performance will be examined under coherent and individual eavesdropping attacks |
| Perumal [28] | √ | √ | × | Quantum channels | ➢ Computational overhead |
| Gong [33] | √ | √ | × | Quantum k-means | ➢ The regular clients have small quantum machines that establish communication with the quantum cloud server |
| Azzaoui [34] | √ | √ | × | quantum cloud as a service | ➢ Performance must be examined under real-time medical data to determine the effective quantum error rate |
| Chen [35] | √ | √ | × | AQ-ABS with IPFS | ➢ Robust to protect the EMRs and resist the various quantum attacks. |
| Mirtskhulava [36] | √ | √ | √ | Post-quantum digital signature | ➢ The performance of the scheme needs to be examined for various quantum attacks. |
| Abdu [42] | √ | √ | × | Grid-based blockchain | ➢ Optimizes the transaction query processing of surgical process records. |

Table 2 depicts a survey of several healthcare quantum computing paradigms, with their use defined as a critical feature. According to Table 2, the existing research focuses on the security and privacy of healthcare data. However, there is a necessity for heterogeneous healthcare device-based implementation in healthcare since these devices have a substantial effect on healthcare service delivery processes.

The significant challenges from state-of-the art are

- The present study of quantum healthcare models' security is insufficient to address the IoT-based security and privacy challenges in medical data sharing using the quantum-blockchain approaches. However, current models are designed for a specific objective.
- The performance analysis of several attacks, such as post-quantum attacks, must be examined in blockchain-based approaches that enable security benefits.
- The practical-oriented quantum blockchain models are essential to provide security and privacy for the healthcare 4.0 industry.

## 4. Software tools and languages

The domain of quantum computing, quantum simulations, and quantum-based software is still earlier stages. Consequently, quantum software tools have been rapidly developed, with several software packages currently accessible from various organizations, including Google, IBM, Microsoft, D-wave, Rigetti, Xanadu, Intel, and Oxford.

However, these tools are still at a rudimentary level, like assembly and high-level languages—the quantum programming tools developed by leveraging C, C++, python, and java. Thus, the usage of QC is increasing exponentially, and organizations are also developing various simulation tools depending on the applications with their comfort languages. The list of the simulation tools and languages used in that particular tool is shown in Table 3.

Table 3: Quantum computing simulation tools [43]

| **Simulator Name** | **Programming Language** | **Web resource** | **status** |
|---|---|---|---|
| Quantum Network Computing | Python | https://sourceforge.net/projects/qnc/files/ | Ineffective |
| qC++ | C++ | https://sourceforge.net/projects/qcplusplus/ | Available |
| Quack! | MATLAB | https://peterrohde.org/introducing-quack/ | Excellence |
| Quantum Information Programs | Mathematica | http://quantum.phys.cmu.edu/QPM/ | Available |
| Feynman | Maple | http://cpc.cs.qub.ac.uk/summaries/ADWE | Available |
| Qubiter | C++ | http://www.ar-tiste.com/qubiter.html | Available |
| QDENSITY | Mathematica | http://www.pitt.edu/~tabakin/QDENSITY/ | Available |
| Linear Al | Mathematica | http://linearal.sourceforge.net/ | Available |
| Quantavo | Maple | http://www3.imperial.ac.uk/quantuminformation/research/downloads | Available |
| Qubit4Matlab | MATLAB | http://bird.szfki.kfki.hu/~toth/qubit4matlab.html | Available |
| Quantum.NET | .Net | https://github.com/phbaudin/quantum-computing | Available |
| Quantum-Octave | GNU Octave and MATLAB | https://github.com/iitis/quantum-octave | Available |
| QuCoSi | C++ | https://sourceforge.net/projects/qucosi/ | Available |
| Bloch Sphere | Java | http://www.ece.uc.edu/~mcahay/blochsphere/ | Completed |
| Quantum eXpress | Java | https://www.componentcontrol.com/solutions/products/quantum-control-family/quantum-express | Available |
| Quantum Circuit | JavaScript | https://www.npmjs.com/package/quantum-circuit | Available |
| QuSAnn | Java | http://www.ar-tiste.com/qusann.html | Available |
| QIO | Qio + Haskell | http://hackage.haskell.org/package/QIO | Available |
| QLib | MATLAB | http://www.tau.ac.il/~quantum/qlib/qlib.html | Available |
| Eqcs | C++ | http://home.snafu.de/pbelkner/eqcs/ | Available |
| Squankum | Java | https://github.com/jeffwass/Squankum | Available |
| TRQS | Mathematica | http://www.iitis.pl/~miszczak/trqs | Available |
| QI | Mathematica | https://github.com/iitis/qi | Available |
| Qitensor | Python | https://qiskit.org/ | Available |
| QuaEC | Python | http://www.cgranade.com/python-quaec/ | Available |
| qMIPS101 | Java | http://institucional.us.es/qmipsmaster/ | Available |
| QuanSuite | Java | http://www.ar-tiste.com/QuanSuite.html | Available |
| QCAD | Xml/Latex | http://qcad.sourceforge.jp/ | Available |
| M-fun | MATLAB/Octave | http://www.ar-tiste.com/m-fun/m-fun-index.html | Available |
| drqubit | MATLAB/Octave | http://www.dr-qubit.org/matlab.php | Available |
| sqct | C++ | https://github.com/vadym-kl/sqct | Available |
| QWalk | C | http://www.cos.ufrj.br/~franklin/qwalk/ | Available |
| jsqis | JavaScript | https://github.com/garrison/jsqis | Available |
| Quipper | Haskell | http://www.mathstat.dal.ca/~selinger/quipper/ | Available |
| QuTip | Python | https://qutip.org/ | Available |
| Quantum++ | C++ | https://github.com/softwareQinc/qpp | Available |
| QuIDE | .Net | http://www.quide.eu/, https://bitbucket.org/quide/quide | Available |
| Quantum | Mathematica | https://homepage.cem.itesm.mx/lgomez/quantum/index.htm | Available |
| QETLAB | MATLAB | http://www.qetlab.com | Available |
| LIQUiD | F# | https://github.com/msr-quarc/Liquid | Available |
| QC-Lib | Python | https://github.com/rnowotniak/qclib | Available |
| SimQubit | C++ | http://sourceforge.net/projects/simqubit/ | Available |
| QuantumUtils | Mathematica | https://github.com/QuantumUtils/quantum-utils-mathematica | Available |
| staq | OpenQASM | https://github.com/softwareqinc/staq | Available |
| ScaffCC | Scaffold | https://github.com/epiqc/ScaffCC | Available |
| Qrack | C++ | https://vm6502q.readthedocs.io | Available |
| QX Simulator | Quantum code | http://quantum-studio.net/ | Available |
| Quintuple | Python | https://github.com/corbett/QuantumComputing?utm_source=catalyzex.com | Available |
| quil | Python | https://www.rigetti.com | Available |
| Quirk | Online tool | https://github.com/Strilanc/Quirk/ | Available |
| Quantum optics.jl | Julia | https://qojulia.org/ | Available |
| Yao. jl | Julia | https://github.com/QuantumBFS/Yao.jl | Available |
| ProjectQ | Python | https://projectq.ch/ | Available |
| Qiskit | Python | https://quantum-computing.ibm.com | Available |
| OpenQASM | QASM | https://github.com/openqasm/openqasm | Available |
| QCGPU | Rust & OpenCL | https://github.com/libtangle/qcgpu?utm_source=catalyzex.com | Available |
| Cirq | Python | https://github.com/quantumlib/Cirq | Available |
| Microsoft Quantum Development Kit | Q# | https://azure.microsoft.com/en-in/resources/development-kit/quantum-computing/ | Available |
| PennyLane | Python | https://github.com/PennyLaneAI/pennylane?utm_source=catalyzex.com | Available |
| QWIRE | Coq | https://github.com/inQWIRE/QWIRE | Available |
| DWave | Python | https://www.dwavesys.com/ | Available |
| QMDD | C++ | http://www.informatik.uni-bremen.de/agra/eng/qmdd.php | Completed |
| Quantum Fog | Graphics programming | http://www.ar-tiste.com/qfog.html | Available |
| Q-Kit | Java | https://sites.google.com/view/quantum-kit/home | Available |
| QSWalk.jl | Julia | https://github.com/QuantumWalks/QSWalk.jl | Available |
| Strawberry | Python | https://pennylane.ai | Available |
| Quantum Computing Playground | Online tool/ Qscript | https://github.com/gwroblew/Quantum-Computing-Playground | Available |
| QuEST | C++ | https://quest.qtechtheory.org | Available |
| Quantum Programming Studio | Java script | https://quantum-circuit.com | Available |
| QCircuits | Python | http://www.awebb.info/qcircuits/index.html | Available |
| IONQ | Python | https://ionq.com/ | Available |
| JKQ | Shell | https://github.com/iic-jku/jkq | Available |
| qHiPSTER | C++ | https://github.com/iqusoft/intel-qs | Available |
| Psitrum | MATLAB | https://github.com/moghadeer/Psitrum | Available |
| Munich Quantum Toolkit (MQT) | OpenQASM | https://github.com/cda-tum/ddsim | Available |
| Quantum Walks | Online tool | http://walk.to/quantum | Available |

According to Table 3, the quantum simulators are available in numerous languages, and their web resources are also listed. The table3 consists of parameters like simulator name, programming language, web resource, and status. The status criteria indicate whether the simulation tool is in use. Nonetheless, some of the tools are currently not working, which are shown as inactive. Some of the tools usages are uncertain (No idea whether tools are working or not) are specified as unknown, and the tools are presently working in various applications development indicated as available tools. Moreover, the available tools are application-specific tools. However, hybrid tools are essential, which is an open research challenge.

Although, the programming languages are leveraged in various simulators based on the developer's interest. Hence the programming languages are employed in most of the simulation tool that is illustrated in Fig.3. According to Fig.3, the languages which are used in the quantum simulator are Python, C++, Matlab, Mathematica, C, JavaScript, Java, Q#, Maple, Julia, Haskell, Perl, Open QASM, GNU Octave, CaML, .Net, Qscript, QASM. Hence, the observation from the Fig.3 is that most of the simulators were initially developed based on the C and C++ languages and later moved to the Matlab and Mathematica tools. At present most of the tools are created through Python or C++.

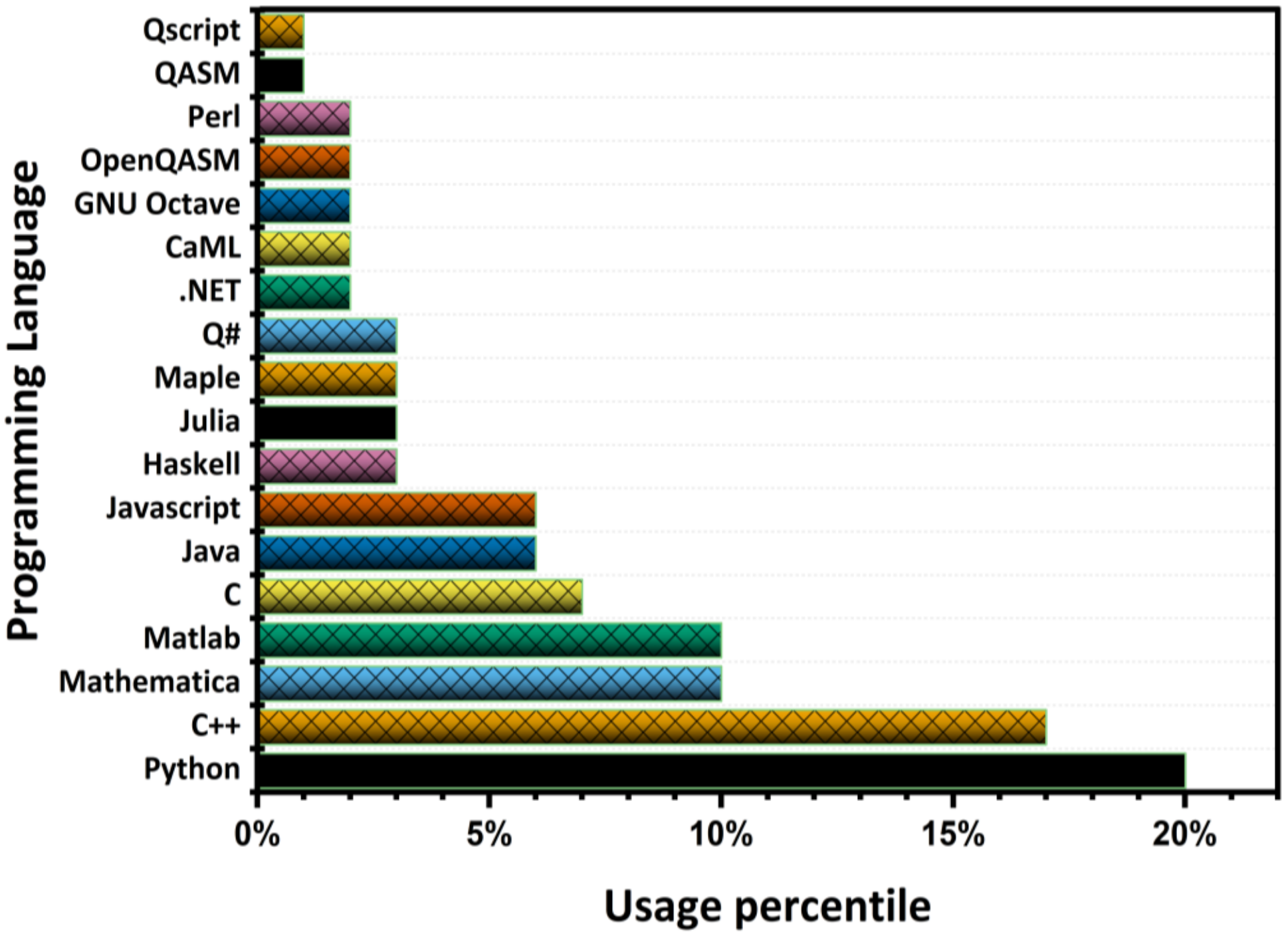

Fig. 3 The percentile usage of programming languages by quantum simulators

## 5. Open issues and future research direction

The critical research issues/problems, challenges, and possible futuristic directions are briefly discussed [15], [44]–[47].

- **Classical healthcare data loading to quantum circuits:** To integrate healthcare-based quantum data with blockchain, classical healthcare data need to be loaded over quantum circuits. The classical to quantum healthcare data loading approaches should be cost-effective. Presently, no focus is drawn towards this direction. Thus, this initiative may be taken up in future.

- **Quantum Blockchain Healthcare Data Structure Presentation**: Presently, there is no standardization to formulate a healthcare data structure using a quantum computing approach. Thus, standardization of the presentation of data in quantum blockchain for efficient implementation is a major challenge for the future.

- **Small-scale Quantum drones for healthcare:** The feasibility to integrate quantum drones with healthcare to improve surveillance and automate the healthcare monitoring system in a secure environment with the least human effort. Secure data communication using quantum computing techniques over drones will improve the data transmission speed and security as well. Thus, theoretical and practical aspects of this study can be taken up in future.

- **High computation and data analysis integration:** QC provides high computational processing that enables personalized services, effective diagnostics, and prognostics. Therefore, leveraging machine learning (ML) and deep learning (DL), the healthcare paradigm data analysis enables descriptive, predictive, diagnostic, and prescriptive analytics.

- **AI/ML for Drug and other Healthcare analysis**: QC provides more opportunities for researchers to conduct artificial Intelligence (AI), ML, and DL operations faster and enhances the performance of the decision support systems in healthcare applications, including Diagnosis pricing analysis, Molecular Simulation, Drug development and recovery, Medical imaging, and recommendation systems.

- **Advanced cryptography and healthcare record security**: Conventional cryptographic algorithms are still suffering from various attacks. Thus, it is daunting for traditional computing systems. Therefore, QC with ML helps in developing breakthroughs in conventional computing. Quantum cryptography, on the other hand, offers more effective methods for protecting data against security attacks on EHR and surgical records.

- **Blockchain and Quantum Blockchain for Healthcare**: The security issues like confidentiality, integrity, trustworthiness, and immutability are provided by blockchain that is collaborated with qubit processing then; it offers high resistance to various attacks like Eclipse, Sybil, Timejacking and Majority (51%) attacks. .

- **Advanced quantum services and pricing strategies:** The resource allocation for quantum service subscribers requires a platform to pay charges per the exploited services and determine the prices. Therefore quantum marketplace development opens the opportunities to develop distributed quantum services strategies and pricing models in financial models, services distributed, and quantum resource distribution.

- **Quantum Computing and Mathematical operations for healthcare**: Mathematical operations and their usages in the healthcare section are yet not explored in detail. For example, quantum cellular automata, quantum dot cellular automata, quantum gates and circuits for short and long-distance secure communication, and many more are yet not explored in detail. We wish to take up these challenges in detail in the future.

- **Quantum computing-based attack analysis for electronic medical records/data**: Medical data protection using blockchain functionalities needs to be examined through QC threats.

## 6. Conclusion

Computing on a quantum level has the potential to result in quantum leaps, which might have far-reaching repercussions for the medical community. This method has the potential to be used in the production of novel medications, the adaptation of treatment strategies to the unique needs of individual patients, and the acceleration of the DNA sequencing process. In order to guarantee that quantum computing is functional in the real world, we investigated the present state of quantum healthcare research and compiled a brief list of the essential requirements that need to be satisfied. In addition, the security of the healthcare delivery system is being investigated in connection with the prospect of using quantum computing. This quantum cryptography research intends to investigate the safety of cryptographic techniques that have previously been examined from the perspective of the healthcare business to identify any potential flaws that may exist.

**Ethics approval and consent to participate**

Not applicable

**Consent for publication**

Not applicable

**Availability of data and materials**

There is no data and material available for this article.

**Conflict of interests**

The authors listed above certify that they have NO affiliations with or involvement in any organization or entity with any financial or non-financial interest (such as personal or professional relationships) affiliations, knowledge or beliefs) in the subject matter or materials discussed in this manuscript.

**Funding:**

There is no funding source available for this article.

## References

[1] X.-M. Hu et al.,"Long-distance entanglement purification for quantum communication," Phys. Rev. Lett., vol. 126, no. 1, p. 010503, 2021, doi: 10.1103/PhysRevLett.126.010503.

[2] T. Jones, A. Brown, I. Bush, and S. C. Benjamin, "QuEST and high-performance simulation of quantum computers," Sci. Rep., vol. 9, no. 1, p. 10736, 2019, doi: 10.1038/s41598-019-47174-9.

[3] P. Srikanth and A. Kumar, "A trustworthy partner selection for mmog using an improved three valued subjective logic uncertainty trust model," PalArch's j. archaeol. Egypt/Egyptol., vol. 18, no. 4, pp. 5677–5698, 2021, https://archives.palarch.nl/index.php/jae/article/view/7160.

[4] P. Srikanth and A. Kumar, "Copyright protection of 3d anaglyph color image watermarking over thin mobile devices," PalArch's j. archaeol. Egypt/Egyptol., vol. 18, no. 6, pp. 15–36, 2021, https://archives.palarch.nl/index.php/jae/article/view/7600.

[5] R. Pifer, "How close is quantum computing in healthcare?," Healthcare Dive, 19-Jul-2021. [Online]. Available: https://www.healthcaredive.com/news/how-close-quantum-computing-in-healthcare-clinical-trials-payers-providers/600554/. [Accessed: 10-Oct-2022].

[6] A. D. V and K. V, "Enhanced BB84 quantum cryptography protocol for secure communication in wireless body sensor networks for medical applications," Pers. Ubiquitous Comput., pp. 1–11, 2021, doi: 10.1007/s00779-021-01546-z.

[7] M. Bhavin, S. Tanwar, N. Sharma, S. Tyagi, and N. Kumar, "Blockchain and quantum blind signature-based hybrid scheme for healthcare 5.0 applications," J. Inf. Secur. Appl., vol. 56, no. 102673, p. 102673, 2021, doi: 10.1016/j.jisa.2020.102673.

[8] A. Steger, "How the internet of medical things is impacting healthcare," Publisher, 16-Jan-2020. [Online]. Available: https://healthtechmagazine.net/article/2020/01/how-internet-medical-things-impacting-healthcare-perfcon. [Accessed: 10-Oct-2022].

[9] W. Rafique, A. S. Hafid, and S. Cherkaoui, "Complementing IoT services using software-defined information-centric networks: A comprehensive survey," IEEE Internet Things J., pp. 1–1, 2022, doi: 10.36227/techrxiv.17159264.v1.

[10] M. Haghi Kashani, M. Madanipour, M. Nikravan, P. Asghari, and E. Mahdipour, "A systematic review of IoT in healthcare: Applications, techniques, and trends," J. Netw. Comput. Appl., vol. 192, no. 103164, p. 103164, 2021, doi: 10.1016/j.jnca.2021.103164.

[11] T. Niedermaier, T. Gredner, S. Kuznia, B. Schöttker, U. Mons, and H. Brenner, "Vitamin D supplementation to the older adult population in Germany has the cost-saving potential of preventing almost 30 000 cancer deaths per year," Mol. Oncol., vol. 15, no. 8, pp. 1986–1994, 2021,doi: 10.1002/1878-0261.12924

[12] D. E. Newman-Toker et al., "Rate of diagnostic errors and serious misdiagnosis-related harms for major vascular events, infections, and cancers: toward a national incidence estimate using the 'Big Three,'" Diagnosis (Berl), vol. 8, no. 1, pp. 67–84, 2021, doi: 10.1515/dx-2019-0104.

[13] Neeraj Chugh Deepak Kumar Sharma Ravi Singhal Saurabh Jain P. Srikanth Adarsh Kumar Alok Aggarwal, "Blockchain-based Decentralized Application (DApp) Design, Implementation, and Analysis with Healthcare 4.0 Trends," Basic & Clinical Pharmacology & Toxicology, vol. 128, no. 2, 2020, doi: 10.1111/bcpt.13405.

[14] "Exploring quantum computing use cases for healthcare," IBM. [Online]. Available: https://www.ibm.com/thought-leadership/institute-business-value/report/quantum-healthcare. [Accessed: 10-Oct-2022].

[15] S. S. Gill et al., "Quantum computing: A taxonomy, systematic review, and future directions," Softw. Pract. Exp., vol. 52, no. 1, pp. 66–114, 2022, doi: 10.1002/spe.3039.

[16] S. W. Hwang et al., "Symbiosis of semiconductors, AI and quantum computing," in 2020 IEEE International Electron Devices Meeting (IEDM), 2020, doi: 10.1109/IEDM13553.2020.9372061.

[17] C. Choudhury, N. Arul Murugan, and U. D. Priyakumar, "Structure-based drug repurposing: Traditional and advanced AI/ML-aided methods," Drug Discov. Today, vol. 27, no. 7, pp. 1847–1861, 2022, doi: 10.1016/j.drudis.2022.03.006.

[18] X. Yuan et al., "Current and Perspective Diagnostic Techniques for COVID-19," ACS Infect. Dis., vol. 6, no. 8, pp. 1998–2016, Aug. 2020, doi: 10.1021/acsinfecdis.0c00365.

[19] A. Ruffino, T.-Y. Yang, J. Michniewicz, Y. Peng, E. Charbon, and M. F. Gonzalez-Zalba, "A cryo-CMOS chip that integrates silicon quantum dots and multiplexed dispersive readout electronics," Nat. Electron., vol. 5, no. 1, pp. 53–59, 2022,doi: 10.1038/s41928-021-00687-6.

[20] Strategyand pwc, "The future of healthcare has arrived | Strategy& Middle East," 2020. https://www.strategyand.pwc.com/m1/en/reports/2021/the-future-of-healthcare-has-arrived.html [accessed Oct. 14, 2022].

[21] K. K. Brock, "Adaptive radiotherapy: Moving into the future," Semin. Radiat. Oncol., vol. 29, no. 3, pp. 181–184, 2019, doi: 10.1016/j.semradonc.2019.02.011.

[22] X. Zheng, S. Sun, R. R. Mukkamala, R. Vatrapu, and J. Ordieres-Meré, "Accelerating Health Data Sharing: A Solution Based on the Internet of Things and Distributed Ledger Technologies," J. Med. Internet Res., vol. 21, no. 6, p. e13583, Jun. 2019, doi: 10.2196/13583.

[23] J. M. Abduljalil and B. M. Abduljalil, "Epidemiology, genome, and clinical features of the pandemic SARS-CoV-2: a recent view," New Microbes New Infect., vol. 35, p. 100672, May 2020, doi: 10.1016/j.nmni.2020.100672.

[24] F. Pappalardo, G. Russo, F. M. Tshinanu, and M. Viceconti, "In silico clinical trials: concepts and early adoptions," Brief. Bioinform., vol. 20, no. 5, pp. 1699–1708, Sep. 2019, doi: 10.1093/bib/bby043.

[25] R. U. Rasool, H. F. Ahmad, W. Rafique, A. Qayyum, and J. Qadir, "Quantum Computing for Healthcare: A Review," 2022, doi: 10.36227/techrxiv.17198702.v2.

[26] W. Rafique, M. Khan, N. Sarwar, and W. Dou, "A security framework to protect edge supported software-defined internet of things infrastructure," in Lecture Notes of the Institute for Computer Sciences, Social Informatics and Telecommunications Engineering, Cham: Springer International Publishing, 2019, pp. 71–88, doi: 10.1007/978-3-030-30146-0_6.

[27] W. Rafique, M. Khan, X. Zhao, N. Sarwar, and W. Dou, "A blockchain-based framework for information security in intelligent transportation systems," in Communications in Computer and Information Science, Singapore: Springer Singapore, 2020, pp. 53–66, doi: 10.1007/978-981-15-5232-8_6.

[28] A. M. Perumal and E. R. S. Nadar, "Architectural framework and simulation of quantum key optimization techniques in healthcare networks for data security," J. Ambient Intell. Humaniz. Comput., 2022, doi: 10.1007/s12652-020-02393-1.

[29] H. Childs, "Applications of cloud-based quantum computers with cognitive computing algorithms in automated, evidence-based Virginia geriatric healthcare," 2020, doi: https://doi.org/10.25886/f8z9-t868.

[30] S. Datta, "Aptamers for Detection and Diagnostics (ADD): Can mobile systems process optical data from aptamer sensors to identify molecules indicating the presence of SARS-CoV-2 virus? Should healthcare explore aptamers as drugs for prevention as well as its use as adjuvants with antibodies and vaccines," ChemRxiv, 2021, doi: 10.26434/chemrxiv.13102877.v32.

[31] T. Janani and M. Brindha, "A secure medical image transmission scheme aided by quantum representation," J. Inf. Secur. Appl., vol. 59, no. 102832, p. 102832, 2021, doi: 10.1016/j.jisa.2021.102832.

[32] A. A. Abd EL-Latif, B. Abd-El-Atty, E. M. Abou-Nassar, and S. E. Venegas-Andraca, "Controlled alternate quantum walks based privacy-preserving healthcare images in the Internet of Things," Opt. Laser Technol., vol. 124, no. 105942, p. 105942, 2020, doi: 10.1016/j.optlastec.2019.105942.

[33] C. Gong, Z. Dong, A. Gani, and H. Qi, "Quantum k-means algorithm based on a trusted server in quantum cloud computing," Quantum Inf. Process., vol. 20, no. 4, 2021, doi: 10.1007/s11128-021-03071-7.

[34] A. EL Azzaoui, P. K. Sharma, and J. H. Park, "Blockchain-based delegated Quantum Cloud architecture for medical big data security," J. Netw. Comput. Appl., vol. 198, no. 103304, p. 103304, 2022, doi: 10.1016/j.jnca.2021.103304

[35] X. Chen, S. Xu, T. Qin, Y. Cui, S. Gao, and W. Kong, "AQ–ABS: Anti-Quantum Attribute-based Signature for EMRs Sharing with Blockchain," in 2022 IEEE Wireless Communications and Networking Conference (WCNC), Apr. 2022, pp. 1176–1181. doi: 10.1109/WCNC51071.2022.9771830.

[36] L. Mirtskhulava, M. Iavich, M. Razmadze, and N. Gulua, "Securing Medical Data in 5G and 6G via Multichain Blockchain Technology using Post-Quantum Signatures," in 2021 IEEE International Conference on Information and Telecommunication Technologies and Radio Electronics (UkrMiCo), Nov. 2021, pp. 72–75. doi: 10.1109/UkrMiCo52950.2021.9716595.

[37] H. B. Mahajan et al., "Integration of Healthcare 4.0 and blockchain into secure cloud-based electronic health records systems," Appl. Nanosci., Feb. 2022, doi: 10.1007/s13204-021-02164-0.

[38] T. M. Fernández-Caramès and P. Fraga-Lamas, "Towards Post-Quantum Blockchain: A Review on Blockchain Cryptography Resistant to Quantum Computing Attacks," IEEE Access, vol. 8, pp. 21091–21116, 2020, doi: 10.1109/ACCESS.2020.2968985.

[39] D. S. Gupta, A. Karati, W. Saad, and D. B. da Costa, "Quantum-Defended Blockchain-Assisted Data Authentication Protocol for Internet of Vehicles," IEEE Trans. Veh. Technol., vol. 71, no. 3, pp. 3255–3266, Mar. 2022, doi: 10.1109/TVT.2022.3144785.

[40] Z. Cai, J. Qu, P. Liu, and J. Yu, "A Blockchain Smart Contract Based on Light-Weighted Quantum Blind Signature," IEEE Access, vol. 7, pp. 138657–138668, 2019, doi: 10.1109/ACCESS.2019.2941153.

[41] A. K. Singh, A. Anand, Z. Lv, H. Ko, and A. Mohan, "A Survey on Healthcare Data: A Security Perspective," ACM Trans. Multimed. Comput. Commun. Appl., vol. 17, no. 2s, p. 59:1-59:26, May 2021, doi: 10.1145/3422816.

[42] N. A. Ali Abdu and Z. Wang, "Grid-based blockchain framework for information management in surgical process proceedings," J. Intell. Fuzzy Syst., vol. 43, no. 4, pp. 5325–5335, Jan. 2022, doi: 10.3233/JIFS-213414.

[43] "List of QC simulators," Quantiki.org. [Online]. Available: https://quantiki.org/wiki/list-qc-simulators. [Accessed: 10-Oct-2022].

[44] A. Kumar et al., “Survey of Promising Technologies for Quantum Drones and Networks,” IEEE Access, vol. 9, pp. 125868–125911, 2021, doi: 10.1109/ACCESS.2021.3109816.

[45] A. Kumar, C. Ottaviani, S. S. Gill, and R. Buyya, “Securing the future internet of things with post-quantum cryptography,” Secur. Priv., vol. 5, no. 2, Mar. 2022, doi: 10.1002/spy2.200.

[46] A. Kumar, D. Augusto de Jesus Pacheco, K. Kaushik, and J. J. P. C. Rodrigues, “Futuristic view of the Internet of Quantum Drones: Review, challenges and research agenda,” Veh. Commun., vol. 36, p. 100487, Aug. 2022, doi: 10.1016/j.vehcom.2022.100487.

[47] K. Kaushik and A. Kumar, “Demystifying quantum blockchain for healthcare,” Secur. Priv., Oct. 2022, doi: 10.1002/spy2.284.